\documentclass[12pt]{article}
\usepackage[english,german,french,polish]{babel}
\usepackage[T1]{fontenc}
\usepackage{amsfonts}

\begin{document}

\selectlanguage{english}

\baselineskip 0.73cm
\topmargin -0.4in
\oddsidemargin -0.1in

\let\ni=\noindent

\renewcommand{\thefootnote}{\fnsymbol{footnote}}

\newcommand{\SM}{Standard Model }

\newcommand{\SMo}{Standard-Model }

\pagestyle {plain}

\setcounter{page}{1}



~~~~~~
\pagestyle{empty}

\begin{flushright}
IFT-- 10/2
\end{flushright}

\vspace{0.4cm}

{\large\centerline{\bf Scattering of nucleons on cold-dark-matter particles}}

{\large\centerline{\bf through photonic portal}}

\vspace{0.5cm}

{\centerline {\sc Wojciech Kr\'{o}likowski}}

\vspace{0.3cm}

{\centerline {\it Institute of Theoretical Physics, University of Warsaw}}

{\centerline {\it Ho\.{z}a 69, 00--681 Warszawa, ~Poland}}

\vspace{0.6cm}

{\centerline{\bf Abstract}}

\vspace{0.2cm}
  
\begin{small}


In the model of hidden sector proposed recently, protons and neutrons scatter differently on cold-dark-matter 
particles. First, we summarize briefly our model based on a mechanism of "photonic portal"~between hidden and
\SMo sectors of the Universe. Then, we calculate in an elementary way the differential cross-sections for scattering of protons and neutrons on sterile Dirac fermions ("\,$\!$sterinos") playing the role of cold-dark
-matter particles. They interact with nucleons through the photonic portal in a somewhat involved but natural 
manner. Due to this portal, the differential cross-section for protons displays a Coulomb-like forward singularity.
 
\vspace{0.6cm}

\ni PACS numbers: 14.80.-j , 04.50.+h , 95.35.+d 

\vspace{0.6cm}

\ni April 2010

 
\end{small}

\vfill\eject

\pagestyle {plain}

\setcounter{page}{1}

\vspace{0.5cm}

\ni {\bf 1. Introduction}

\vspace{0.4cm} 

In this note, we calculate the differential cross-sections for scattering of protons and neutrons on cold-dark-matter particles in the model [1,2], where cold dark matter consists of stable sterile Dirac fermions ("\,$\!$sterinos"). They interact weakly in a hidden sector together with unstable sterile spin-0 bosons ("\,$\!$sterons") {\it via} sterile non-gauge mediating bosons ("$A$ bosons")  described by an antisymmetric-tensor field (of dimension one). In this interaction, sterons are paired with \SMo photons, so that a new weak interaction Lagrangian in the hidden sector reads:

\begin{equation}
- \frac{1}{2} \sqrt{f}\left(\varphi F_{\mu \nu} + \zeta \bar\psi \sigma_{\mu \nu} \psi \right) A^{\mu \nu}
\end{equation}

\ni with $\sqrt{f}$ and $\sqrt{\!f}\,\zeta$ denoting two dimensionless small coupling constants, while $F_{\mu \nu} = \partial_\mu A_\nu - \partial_\nu A_\mu$ is the \SMo electromagnetic field (of dimension two). Here, we assume that $\varphi = <\!\!\varphi\!\!>_{\rm vac}\! + \,\varphi_{\rm ph}$, where  $<\!\!\varphi\!\!>_{\rm vac} \neq 0$ is a spontaneously nonzero vacuum expectation value of the steron field. The coupling (1) of photons to the hidden sector has been called "photonic portal"\, (to hidden sector). It is an alternative to the popular "Higgs portal"\, (to hidden sector) [3]. 

Together with the $A$-boson kinetic and \SMo electromagnetic Lagrangians, the new interaction Lagrangian (1) leads to the following field equations for $F_{\mu \nu}$ and $A_{\mu \nu}$:

\begin{equation}
\partial^\nu (F_{\mu \nu} +  \sqrt{\!f}\, \varphi A_{\mu \nu}) = -j_\mu \;\;,\;\; F_{\mu \nu} = \partial_\mu A_\nu - \partial_\nu A_\mu 
\end{equation}

\ni and 

\begin{equation}
(\Box - M^2)A_{\mu \nu} = - \sqrt{f} (\varphi F_{\mu \nu} + \zeta \bar\psi \sigma_{\mu \nu} \psi) \,,
\end{equation}

\ni where $j_\mu$ stands for the \SMo electric current and $M$ denotes a mass scale of $A$ bosons, expected typically to be large.

The field equations (2) are Maxwell's equations modified due to the presence of hidden sector. Such a modification has a magnetic character, since the hidden-sector contribution to the total electric source current

\begin{equation}
j_\mu + \partial^\nu ( \sqrt{\!f}\, \varphi A_{\mu \nu})
\end{equation}

\ni for the electromagnetic field $A_{\mu }$ is here a four-divergence not contributing to the total electric charge 

\begin{equation}
\int d^3x[j_0 + \partial^k ( \sqrt{\!f}\, \varphi A_{0 k})] = \int d^3x j_0 = Q\,.
\end{equation}

In particular, the nonzero vacuum expectation value $<\!\!\varphi\!\!>_{\rm vac} \neq 0$ generates spontaneously an effective sterino magnetic interaction

\begin{equation}
- \frac{1}{2} \mu_\psi (\bar\psi \sigma_{\mu \nu} \psi) F^{\mu \nu}\,,  
\end{equation}

\ni though sterinos are electrically neutral. Here,

\begin{equation}
\mu_\psi = \frac{f \zeta<\!\!\varphi\!\!>_{\rm vac}}{M^2}  
\end{equation}

\ni is a resulting sterino magnetic moment. The interaction (6) follows as a part of the effective interaction

\begin{equation}
  -\frac{1}{4} \frac{f}{M^2}\left(\varphi F_{\mu \nu} + \zeta \bar\psi \sigma_{\mu \nu} \psi \right) \left(\varphi F^{\mu \nu} + \zeta \bar\psi \sigma^{\mu \nu} \psi \right) \,,
\end{equation}

\ni when the low-momentum-transfer approximation

\begin{equation}
 A^{\mu \nu} \simeq \frac{\sqrt{f\,}}{M^2}\left(\varphi F^{\mu \nu} + \zeta \bar\psi \sigma^{\mu \nu} \psi \right)\,
\end{equation}

\ni implied by Eq. (3) is used in the interaction (1) with $\varphi = <\!\!\varphi\!\!>_{\rm vac}\! + \,\varphi_{\rm ph}$.{\footnote{To avoid this approximation, one may apply in the effective interaction (8) in place of $(M^2)^{-1}$ the Green's function $(M^2- \Box)^{-1}$ inserted between both brackets (~), leading to a nonlocal effective interaction operator.}} The quadratic form (8) is multiplied by an extra factor $1/2$ in order to work as an effective Lagrangian.

The \SMo Dirac fermions $f$ ({\it e.g.} $f = e^-, \mu^-, p, n$) interact electromagnetically according to the following interaction Lagrangian:

\begin{equation}
-\bar\psi_f (e_f\, \gamma^\mu A_\mu + \frac{1}{2}\delta \mu_f\, \sigma^{\mu \nu} F_{\mu \nu}) \,\psi_f \,, 
\end{equation}

\ni where

\begin{equation}
\delta \mu_f = \frac{g_f}{2} \frac{e}{2 m_f} - \frac{e_f}{2m_f}\;\;,\;\; \frac{e^2}{4\pi} = \alpha \simeq \frac{1}{137} \,.
\end{equation}

\ni In particular, for protons and neutrons (treated as pointlike particles)

\vspace{-0.1cm}

\begin{equation}
e_p = e \;\;,\;\; \delta \mu_p \simeq 1.79 \,\frac{e}{2m_p} \;\;;\;\; e_n = 0 \;\;,\;\;\delta \mu_n \simeq -1.91 \,\frac{e}{2m_n} \,.
\end{equation}

\vspace{-0.1cm}

\ni In Eq. (10), we have

\vspace{-0.1cm}

\begin{equation}
\frac{1}{2}\sigma_{\mu \nu} F^{\mu \nu} = i \,\vec{\alpha}\cdot\vec{E} - \vec{\sigma}\cdot\vec{B} \,,
\end{equation}

\vspace{-0.1cm}

\ni where $\vec{\alpha} = \beta \,\vec{\gamma} = \gamma_5 \,\vec{\sigma}$ with $(\gamma^\mu) = (\beta, 
\vec{\gamma})  = (\beta, \beta \vec{\alpha})$. The total magnetic moment of an $f$ fermion is

\vspace{-0.1cm}

\begin{equation}
{\mu}_f = \frac{e_f}{2m_f}+ \delta\mu_f =  \frac{g_f}{2} \frac{e}{2m_f}\;\;,\;\; \frac{e^2}{4\pi} = \alpha \simeq \frac{1}{137}\,, 
\end{equation}

\vspace{-0.1cm}

\ni while

\vspace{-0.1cm}

\begin{equation}
\vec{\mu}_f =  \mu_f \,\vec{\sigma} =  \frac{g_f}{2} \frac{e}{2m_f} \vec{\sigma} \,.
\end{equation}

\ni In particular, $\mu_p \simeq 2.79(e/2m_p)$ and $\mu_n \simeq -1.91(e/2m_n)$.

\vspace{0.3cm}  

{\bf 2. Scattering of nucleons on sterinos}

\vspace{0.3cm}

The current direct detection experiments for cold dark matter (for theoretical aspects {\it cf.} [4,5]) aim to observe recoils of Earth's detector nuclei scattered on dark-matter particles during the travel of the Earth through the Galactic halo. From the point of view of nuclear physics, this is a nonrelativistic scattering problem for nucleons bound within nuclei moving in a medium of particle cold dark matter and weakly interacting with it by a postulated new coupling. In this note, we consider this scattering using our model of hidden sector, where its coupling to \SMo sector is provided by the photonic portal described in Introduction. We study the free  nucleon scattering rather than the nuclear scattering on cold-dark-matter particles, in order to look as close as possible at an elementary mechanism. Of couse, in realistic detection experiments nucleons are bound within detector nuclei.

In our model, pointlike protons interact with sterinos (constituting the cold dark matter) {\it via} the part of weak-interaction Lagrangian (1) involving $<\!\!\varphi\!\!>_{\rm vac} \neq 0$,

\begin{equation}
- \frac{1}{2} \sqrt{f}\left(<\!\!\varphi\!\!>_{\rm vac} F_{\mu \nu} + \zeta \bar\psi \sigma_{\mu \nu} \psi \right) A^{\mu \nu}\,,
\end{equation}

\ni together with both parts of electromagnetic interaction Lagrangian (10), 

\begin{equation}
-e \,\bar\psi_p \gamma^\mu \psi_p\, A_\mu \;\;\;{\rm and}\;\;\;- \frac{1}{2}\delta \mu_p \,\bar\psi_p \sigma^{\mu \nu} \psi_p \,F_{\mu \nu} \;,
\end{equation}

\ni while pointlike neutrons do that {\it via} the part (16) together with the second part of Eq. (10) only,

\begin{equation}
- \frac{1}{2}\delta \mu_n \,\bar\psi_n \sigma^{\mu \nu} \psi_n\, F_{\mu \nu} \;.
\end{equation}

The interaction (16), when collaborating with the first coupling (17), leads for the scattering $p \psi \rightarrow p' \psi'$ to the following $S$-matrix element (in the obvious notation):

\begin{eqnarray}
S_{\rm el}(p \psi \rightarrow p' \psi') & = & -i\frac{1}{4}\frac{e f \zeta \!<\!\!\varphi\!\!>_{\rm vac}}{k^2(k^2-M^2)} \left[\frac{1}{(2\pi)^{12}} \frac{m^2_p\,m^2_\psi}{E'_p E_p E'_\psi E_\psi} \right]^{1/2}\!\! \left(2\pi \right)^4 \delta^4 \left( p'_p \!+\! p'_{\psi} \!-\! p_p \!-\! p_{\psi}\right) \nonumber \\
 & \times & \left[ \bar{u}'_p(p'_p) \gamma^\mu u_p(p_p)\right] 2 \left[ \bar{u}'_\psi(p'_\psi) \sigma_{\mu\,\nu}(p'_\psi-p_\psi)^\nu u_\psi(p_\psi) \right] \,,
\end{eqnarray}

\vspace{0.2cm}

\ni where $k = p'_p-p_p = p_\psi-p'_\psi$ and, in the next step, the Gordon identity will be used:

\begin{equation}
i\, \bar{u}'(p') \sigma_{\mu\nu}(p' - p)^\nu u(p) = \bar{u}'(p') \left[2m\gamma_\mu -(p'+p)_\mu\right] u(p)\,.
\end{equation}

\ni In Eq. $\!\!$(19), the factor 2 standing at $-\bar{u}'_\psi\,\sigma_{\mu\,\nu}\,k^\nu\, u_\psi$ comes out from the prod\-uct $[(\bar{\psi}_p \gamma^\mu \psi_p)A_\mu](x)\,[F^{\rho \sigma}(\bar{\psi}\sigma_{\rho \sigma}\psi)](x')$ giving $(1/k^2)(\bar{u}'_p \gamma^\mu u_p)(g^{\mu \sigma}k^\rho - g^{\mu \rho} k^\sigma)(\bar{u}'_\psi \sigma_{\rho \sigma} u_\psi) = (1/k^2)(\bar{u}'_p \gamma^\mu u_p) 2 (-\bar{u}'_\psi \sigma_{\mu \nu} k^\nu u_\psi)$. 

The collaboration of interaction (16) with the second coupling (17) or with the coupling (18) provides for the scattering $f \psi\, \rightarrow f' \,\psi'$ the following $S$-matrix element:

\begin{eqnarray}
S_{\rm mag}(f \psi\! \rightarrow \!f' \psi') \!\!&\!\! = \!\!&\!\!-i\,\frac{1}{8}\, \frac{\delta \mu_f\,f \zeta\!<\!\!\varphi\!\!>\!_{\rm vac}}{k^2(k^2-M^2)}\! \left[\!\frac{1}{(2\pi)^{12}}\, \frac{m^2_f m^2_\psi}{E'_f E_f\, E'_\psi E_\psi}\right]^{\!1/2}\!\!\!\!\! \left(2\pi \right)^4 \delta^4 \left(p'_f \!+\! p'_\psi \!-\! p_f \!-\! p_\psi\!\right) \nonumber \\
 \!\!&\!\! \times\!\! &\!\! 2\left[\bar{u}'_f(p'_f) \sigma^{\mu \lambda}(p'_f\!-\!p_f)_\lambda u_f(p_f)\!\right] 2 \left[\bar{u}'_\psi(p'_\psi) \sigma_{\mu \nu}(p'_\psi\!-\!p_\psi)^\nu u_\psi(p_\psi)\!\right]  
\end{eqnarray}

\vspace{0.2cm}

\ni with $f = p$ or $f = n$. 

Since the fully differential cross-section corresponding to the $S$-matrix element $S(f \psi\, \rightarrow f' \,\psi')$ is defined as

\begin{equation}
\frac{d^6 \sigma(f \psi\, \rightarrow f' \,\psi')}{d^3 \vec{p}\,'_f d^3 \vec{p}\,'_\psi} = \frac{(2\pi)^6}{v_{\rm rel}}\sum_{u'_f u'_\psi} \,\frac{1}{4} \sum_{u_f u_\psi}\,\frac{|S(f \psi\, \rightarrow f' \,\psi')|^2}{(2\pi)^4 \delta^4(0)}\,, \end{equation}

\ni we get in the case of $S$-matrix element (19) for protons:

\begin{eqnarray}
\frac{d^6 \sigma_{\rm el}(p \psi\, \rightarrow p' \psi')}{d^3 \vec{p}\,'_p d^3 \vec{p}\,'_\psi} \!\!\!& \!\!=\!\! &\!\!\frac{1}{v_{\rm rel}} \,\frac{1}{16} \left( \frac{e f \zeta<\varphi>_{\rm vac}}{M^2-k^2}\right)^2\!\! \frac{1}{(2\pi)^2}\, \frac{m^2_\psi}{E'_p E_p E'_\psi E_\psi}\, \delta^4\!\left(p'_p \!+\! p'_\psi \!-\!p_p \!-\!p_\psi \right)  \nonumber \\
&\!\!\times\!\! &\!\!\frac{m^2_p}{k^4}\sum_{u'_p u'_\psi}\,\sum_{u_p u_\psi} |(\bar{u}'_p \gamma^\mu u_p)\!\!\left\{\bar{u}'_\psi \!\left[2m_\psi\gamma_\mu \!-\! (p'_\psi \!+\! p_\psi)_\mu\right]\! u_\psi\!\right\}\!|^2 \,. 
\end{eqnarray}

\ni Evaluating traces in Dirac bispinor indices, we obtain in Eq. (23):

\begin{equation} 
\frac{m^2_p}{k^4} \sum_{u'_p u'_\psi}\!\sum_{u_p u_\psi} |\;\;|^2 = 1 - 2\frac{p_p \cdot p_\psi}{m^2_\psi} + 4 
\left[ m^2_p - \frac{(p_p \cdot p_\psi)^2}{m^2_\psi}\right]\frac{1}{k^2}\;,
\end{equation}

\vspace{-0.2cm}

\ni where 

\vspace{-0.2cm}

\begin{equation} 
k^2 = (p'_p-p_p)^2  = 2(m^2_p -p'_p\!\cdot\!p_p) = 2\left(m^2_p - E'_p E_p + |\vec{\,p}'_p||\vec{\,p}_p| \cos\theta'_p\right)
\end{equation}

\ni with $\theta'_p$ being the angle between the directions of $\vec{\,p}'_p$ and $\vec{\,p}_p$.

Finally, using the definition of differential cross-section

\begin{equation}
\frac{d \sigma(f \psi\, \rightarrow f' \psi')}{d \Omega'_f} = \int_{0}^{\infty}{{\vec{p}\,'\!\!}_f}^2 d|{\vec{p}\,'\!\!}_f|\,\int d^3 {\vec{p}\,'\!\!}_\psi \frac{d^6 \sigma(f \psi\, \rightarrow f' \psi')}{d^3 {\vec{p}\,'\!\!}_f\, d^3 {\vec{p}\,'\!\!}_\psi} \;,
\end{equation}

\ni we calculate in the case of $S$-matrix element (19) for protons:

\begin{eqnarray}
\frac{d \sigma_{\rm el}(p \psi\, \rightarrow p' \psi')}{d \Omega'_p} \!\!\! &\!\! =\!\! & \frac{1}{v_{\rm rel}}\,\frac{1}{16} \left( \frac{e f \zeta\!<\!\varphi\!>_{\rm vac}}{M^2 - k^2}\right)^2 \!\frac{1}{(2\pi)^2}\, \frac{m^2_\psi \vec{\,p}'\!^{\;2}_p}{E'_p E_p E'_\psi E_\psi}\, \frac{d|\vec{\,p}'_p|}{d(\!E'_p \!+\! E'_\psi)} \nonumber \\
 \!\! & \!\!\times\!\! & \!\!\left\{1-2\frac{p_p \!\cdot\! p_\psi}{m^2_\psi} + 4\left[ m^2_e - 
\frac{(p_p \cdot p_\psi)^2}{m^2_\psi}\right]\frac{1}{k^2}\right\}\;,
\end{eqnarray}

\ni where $p'_p + p'_\psi = p_p + p_\psi$ and $d \Omega'_p = 2\pi \sin \theta'_p d\theta'_p$.

Further, we calculate this differential cross-section in the centre-of-mass frame, where $\vec{p}_p + \vec{p}_\psi = 0$ and so, $\vec{\,p}'_{p} + \vec{\,p}'_{\psi} = 0$. Then, $|\vec{\,p}'_p| = |\vec{\,p}_p|\,,\,E'_p = E_p$ and $|\vec{\,p}'_\psi| = |\vec{\,p}_\psi|\,,\,E'_\psi = E_\psi$ as well as

\begin{equation}
v_{\rm rel} = \frac{|\vec{p}_p|}{E_p} + \frac{|\vec{p}_\psi|}{E_\psi} = \frac{{E_p} + E_\psi}{E_p E_\psi}\,|\vec{p}_p|\,,\,\frac{d(E'_p+E'_\psi)}{d|\vec{\,p}'_p|} = \frac{|\vec{\,p}'_p|}{E'_p} + \frac{|\vec{\,p}'_\psi|}{E'_\psi} = \frac{E_p + E_\psi}{E_p E_\psi} |\vec{\,p}_p| \,.
\end{equation}

\ni In this frame, Eq. (27) gives

\vspace{-0.3cm}

\begin{eqnarray}
\frac{d \sigma_{\rm el}(p \psi \rightarrow \!p' \psi')}{d \Omega'_p} \!\!\! &\!\! =\!\! & \!\!\frac{1}{16}
\left( \frac{e f \zeta\!<\!\varphi\!>_{\rm vac}}{M^2 - k^2}\right)^2 \!\frac{1}{(2\pi)^2} \,\frac{m^2_\psi}{(E_p + E_\psi)^2} \nonumber \\  \!\!\! & \!\!\times\!\! & 
\!\!\left\{\!1\!-\!\frac{(E_p\!+\!E_\psi)^2\! -\! m^2_p\! -\!m^2_\psi}{m^2_\psi} \!+\! 4\!\left[m^2_p \!-\! \frac{\left((E_p\!+\!E_\psi)^2\! -\! m^2_p -m^2_\psi \right)^2}{4m^2_\psi}\! \right]\frac{1}{k^2}\right\}\,, \nonumber \\
\end{eqnarray}

\vspace{-0.4cm}

\ni where

\vspace{-0.4cm}

\begin{equation} 
k^2 = (p'_p-p_p)^{\!2}  = 2\left(m^2_p - E^2_p + \vec{\,p}^{\,2}_p \cos\theta'_p\right) = -4 \vec{\,p}^{\,2}_p\sin^2 \frac{\theta'_p}{2}\,.
\end{equation}

Since momenta involved in the cold-dark-matter interaction with nuclear detectors are not larger than a few keV, the nonrelativistic approximation $E = \sqrt{\vec{\,p}^2 + m^2} \sim m + \vec{\,p}^2/2m \simeq m$ is adequate. Then, the differential cross-section (29) in the leading nonrelativistic approximation takes the form

\vspace{-0.2cm}

\begin{equation} 
\frac{d \sigma_{\rm el}(p \psi \rightarrow \!p' \psi')}{d \Omega'_p} \! \simeq \!\frac{1}{16} \!\left( \frac{e f \zeta\!<\!\varphi\!>_{\rm vac}}{M^2}\right)^{\!2} \!\!\frac{1}{(2\pi)^2} \,\frac{m^2_\psi}{(m_p \!+\! m_\psi)^2}\!\left[\!1\!-\!2\frac{m_p}{m_\psi}\left(\!1\!-\!\frac{1}{\sin^2 \frac{\theta'_p}{2}}\right)\!\right]. 
\end{equation}

We can see that this differential cross-section for protons gets a Coulomb-like forward singularity provided by our photonic-portal mechanism of interaction between hidden and \SMo sectors of the Universe. This singularity is weaker than in the Rutherford cross-section.

In a similar way for neutrons, using the $S$-matrix element (21) with $f = n$, we get

\vspace{-0.1cm}

\begin{eqnarray}
\frac{d^6\sigma_{\rm mag}(n \psi \rightarrow \!n' \psi')}{d^3 \vec{\,p}'_n d^3 \vec{\,p}'_\psi}\!\! &\!\! =\!\! & \!\frac{1}{v_{\rm rel}} \frac{1}{16}\! \left( \frac{\delta\mu_n f \zeta\!<\!\varphi\!>_{\rm vac}}{M^2 - k^2}\right)^{\!2} 
\!\!\frac{1}{(2\pi)^2} \,\frac{m^2_n m^2_\psi}{E'_n E_n E'_\psi E_\psi}\delta^4(p'_n\!+\!p'_\psi\!-\! 
p_n\!-\!p_\psi) \nonumber \\  \!\! & \!\!\times\!\! & \!\frac{1}{k^4}\sum_{u'_n u'_\psi} \sum_{u_n u_\psi}
|\!\left[\!\bar{u}'_n\, \sigma^{\mu\lambda}(p'_n\!-\!p_n)_{\!\lambda}\, u_n\!\right]
\!\left[\!\bar{u}'_\psi\, \sigma_{\mu \nu}(p'_\psi \!-\! p_\psi)^\nu\, u_\psi\!\right]\!|^2\,, 
\end{eqnarray}

\ni where, in the next step, it is convenient to apply twice the Gordon identity (20). Then, evaluating traces in Dirac bispinor indices, we obtain in Eq. (32)

\vspace{-0.1cm}

\begin{equation} 
\frac{1}{k^4} \sum_{u'_n u'_\psi}\!\sum_{u_n u_\psi} |\;\;|^2 = 4+ 4\frac{(p_n \!\cdot\! p_\psi)^2}{m^2_n m^2_\psi} + \frac{(p_n \!+\! p_\psi)^2}{m^2_n m^2_\psi}k^2 + \frac{1}{4m^2_n m^2_\psi}k^4 \;,
\end{equation}

\ni where 

\begin{equation} 
k^2 = (p'_n-p_n)^2  = 2(m^2_n -p'_n\!\cdot\!p_n) = 2\left(m^2_n - E'_n E_n + |\vec{\,p}'_n||\vec{p}_n| \cos\theta'_n\right)\,.
\end{equation}

\ni Finally, the differential cross-section reads

\begin{eqnarray}
\frac{d\sigma_{\rm mag}(n \psi \rightarrow \!n' \psi')}{d\Omega'_n}\!\! &\!\! =\!\! & \!\frac{1}{v_{\rm rel}} \frac{1}{16} \left( \frac{\delta\mu_n f \zeta\!<\!\varphi\!>_{\rm vac}}{M^2 - k^2}\right)^2 
\!\frac{1}{(2\pi)^2} \,\frac{m^2_n m^2_\psi \vec{\,p}'^2_n}{(E'_n E_n E'_\psi E_\psi)^2}\frac{d |\vec{\,p}'_n|}{d(E'_n+E'_\psi)} \nonumber \\ \!\! & \!\!\times\!\! & \!\left[4+ 4\frac{(p_n \!\cdot\! p_\psi)^2}{m^2_n m^2_\psi} + \frac{(p_n \!+\! p_\psi)^2}{m^2_n m^2_\psi}k^2 + \frac{1}{4m^2_n m^2_\psi}k^4  \right] \,.
\end{eqnarray}

\ni In the centre-of-mass frame, where$\vec{\,p}_n+\vec{\,p}_\psi = 0$ and so $\vec{\,p}'_n+\vec{\,p}'_\psi = 0$, we have $|\vec{\,p}'_n| = |\vec{\,p}_n|$, $E'_n = E_n$ and $|\vec{\,p}'_\psi| = |\vec{\,p}_\psi|$, $E'_\psi = E_\psi$ as well as

\begin{equation}
v_{\rm rel} = \frac{{E_n} + E_\psi}{E_n E_\psi}\,|\vec{p}_n|\,,\,\frac{d(E'_n+E'_\psi)}{d|\vec{\,p}'_n|} = \frac{E_n + E_\psi}{E_n E_\psi} |\vec{\,p}_n| \,.
\end{equation}

\ni In this frame, Eq. (35) implies

\begin{eqnarray}
\frac{d \sigma_{\rm mag}(n \psi \rightarrow \!n' \psi')}{d \Omega'_n} \!\! &\!\! =\!\! & \!\frac{1}{16} \left( \frac{\delta\mu_n f \zeta\!<\!\varphi\!>_{\rm vac}}{M^2 - k^2}\right)^2 
\!\frac{1}{(E_n+E_\psi)^2} \nonumber \\ 
\!\! & \!\!\times\!\! & \!\left\{4+ \frac{[(E_n+E_\psi)^2- m_n^2 -m_\psi^2]^2}{m^2_n m^2_\psi} + \frac{(E_n+E_\psi)^2}{m^2_n m^2_\psi}k^2 + \frac{1}{4m^2_n m^2_\psi}k^4  \right\} \,, \nonumber \\
\end{eqnarray}

\vspace{-0.2cm}

\ni where

\vspace{-0.2cm}

\begin{equation} 
k^2 = (p'_n-p_n)^2  = 2(m^2_n - E_n^2+ \vec{\,p}^2_n\cos \theta'_n) = -4\vec{\,p}^2_n\sin^2 \frac{\theta'_n}{2}\,.
\end{equation}

\ni The differential cross-section (37) in the leading nonrelativistic approximation becomes

\begin{equation} 
\frac{d \sigma_{\rm mag}(n \psi \rightarrow n' \psi')}{d \Omega'_n} \simeq \frac{1}{2} \left( \frac{\delta\mu_n f \zeta<\!\varphi\!>_{\rm vac}}{M^2}\right)^2 \frac{1}{(2\pi)^2} \frac{m^2_n m^2_\psi}{(m_n + m_\psi)^2} \;,
\end{equation}

\ni giving trivially the integral cross-section in this approximation:

\vspace{-0.2cm}

\begin{equation} 
\sigma_{\rm mag}(n \psi \rightarrow n' \psi') \simeq \frac{1}{2\pi} \left( \frac{\delta\mu_n f \zeta<\!\varphi\!>_{\rm vac}}{M^2}\right)^2 \frac{m^2_n m^2_\psi}{(m_n + m_\psi)^2} \;,
\end{equation}

\ni where $\delta\mu_n  = \mu_n = (g_n/2)(e/2 m_n)$ and so,

\vspace{-0.3cm}

\begin{equation} 
\frac{1}{2} \left(\frac{\delta\mu_n f \zeta<\!\varphi\!>_{\rm vac}}{M^2}\right)^2 m^2_n =  \frac{1}{8}\left( \frac{g_n}{2}\right)^2\left(\frac{e f \zeta<\!\varphi\!>_{\rm vac}}{M^2}\right)^2  \,.
\end{equation}

It can be seen from the differential cross-section (37) or (39) for neutrons that, in contrast to the case of protons, it does not get a Coulomb-like forward singularity. The same is true in the case of protons for the differential cross-section $d \sigma_{\rm mag}(p\psi \rightarrow p'\psi')/d\Omega'_p$ which takes the form analogical to (37) or (39) but with $f=n$ replaced by $f=p$, where $\delta\mu_p = \mu_p - e/2m_p = (g_p/2 - 1)(e/2m_p)$.

However, in the case of protons, in their differential cross-section there is still a third term implied by the interference in the expression $|S_{\rm el} + S_{\rm mag}|^2 = |S_{\rm el}|^2 + | S_{\rm mag}|^2 + (S_{\rm el}S^*_{\rm mag} + S_{\rm mag} S^*_{\rm el}$). After an elementary calculation, we obtain in the centre-of-mass frame, where $\vec{\,p}_p + \vec{\,p}_\psi = 0$, the following interference differential cross-section:

\vspace{-0.3cm}

\begin{equation} 
\frac{d \sigma_{\rm elmag}(p \psi \rightarrow p' \psi')}{d \Omega'_p} = \frac{1}{16} \frac{e\, \delta\mu_p\left(f \zeta\!<\!\varphi\!>_{\rm vac}\!\right)^2}{M^4} \,\frac{1}{(2\pi)^2}\, 
\frac{m_p m^2_\psi}{(E_p + E_\psi)^2} \left(8 + \frac{m^2_p+ m^2_\psi}{m^2_p m^2_\psi} k^2\right) \,,
\end{equation}

\vspace{-0.2cm}

\ni where

\vspace{-0.3cm}

\begin{equation} 
k^2 = (p'_p-p_p)^2  = 2\left(m^2_p - E_p^2 + \vec{\,p}^2_p \cos\theta'_p \right) = -4\vec{\,p}'^{\,2}_p \sin{\frac{\theta'}{2}} \,.
\end{equation}

\ni Hence, in the leading nonrelativistic approximation

\vspace{-0.3cm}

\begin{equation} 
\frac{d \sigma_{\rm elmag}(p \psi \rightarrow p' \psi')}{d \Omega'_p} \simeq \frac{1}{2} \frac{e\, \delta\mu_p (f \zeta\!<\!\varphi\!>\!_{\rm vac})^2}{M^4} \,\frac{1}{(2\pi)^2}\,\frac{m_p m^2_\psi} {(m_p+ m_\psi)^2} \,,
\end{equation}

\vspace{-0.2cm}

\ni with

\vspace{-0.3cm}

\begin{equation} 
\frac{1}{2} \frac{e\, \delta\mu_p\left(f \zeta\!<\!\varphi\!>\!_{\rm vac}\!\right)^2}{M^4} \,m_p = \frac{1}{4}\left(\frac{g_p}{2} - 1\right) \left(\frac{e f \zeta\!<\!\varphi\!>_{\rm vac}}{M^2}\!\right)^2 \,,
\end{equation}

The sum of three differential  cross-sections in the case of protons gives in the leading nonrelativistic approximation the following formula (see Eqs. (31), (39) with $f=p$ in place of  $f=n$ and (44)):

\vspace{-0.3cm}
 
\begin{eqnarray} 
\!\!\!\!\left(\frac{d \sigma_{\rm el}}{d \Omega'_p} \!+\! \frac{d \sigma_{\rm mag}}{d \Omega'_p} \!+ \!\frac{d \sigma_{\rm elmag}}{d \Omega'_p}\right)\!\!(p \psi \!\rightarrow\! p' \psi') &\!\! \simeq \!\!& \!\!\frac{1}{16} \left(\frac{e f \zeta\!<\!\!\varphi\!\!>\!_{\rm vac}}{M^2}\!\right)^2\! \frac{1}{(2\pi)^2} \frac{m^2_\psi}{(m_p \!+\!m_\psi)^2} \nonumber \\ \!\!\!\!\!& \!\!\times\!\! & \!\!\!\left[\!1\!-\!\frac{m_p}{m_\psi}\!\left(\!1\!-\!\frac{1}{\sin^2\frac{\theta'_p}{2}}\!\right)\!+\!2\left(\!\frac{g_p}{2}\!-\!1\!\right)^2\!+\!4\left(\!\frac{g_p}{2}\!-\!1\!\right)\!\right]\,. \nonumber \\
\end{eqnarray}

\ni Here, $g_p/2 - 1 \simeq 1.79$, while $g_n/2 \simeq 1.91$  in Eqs. (39) with (41). Of course, the first partial cross-section in Eq. (46) displays a Coulomb-like forward singularity.

Introducing an experimental uncertainty $|k^2|_{\rm min}$ for measurement of small momentum transfers $|k^2| = 4\vec{\,p}^2_p \,\sin^2(\theta'_p/2)$ at fixed $\vec{\,p}^2_p $, we get a minimal angle $\theta_{\rm min}$ defined as $\sin^2(\theta_{\rm min}/2) = |k^2|_{\rm min}/(4\vec{\,p}^2_p)$. Then, denoting the lhs of Eq. (46) by $d \sigma(p \psi \rightarrow p' \psi')/d\Omega'_{p}$ , we integrate trivially over $\theta'_p$ from $\theta_{\rm min}$ to $\pi$ obtaining

\begin{eqnarray} 
\sigma(p \psi \rightarrow p' \psi') & \simeq & \frac{1}{16} 
\left(\frac{e f \zeta\!<\!\!\varphi\!\!>\!_{\rm vac}}{M^2}\right)^2 
\frac{1}{(2\pi)^2} \frac{ m^2_\psi}{(m_p + m_\psi)^2} \nonumber \\ & \times & 
\left\{4\pi\! \left[\!1\!-\!\frac{m_p}{m_\psi} \!+2\left(\!\frac{g_p}{2}-1\!\right)^2\!+4\!\left(\frac{g_p}{2}\!-1\!\right) \!\right]\! +4\frac{m_p}{m_\psi}\ln\frac{4\vec{\,p}^2_p}{|k^2|_{\rm min}}\!\right\}\,.
\end{eqnarray}

\ni Perhaps, a theoretical estimate of minimal $|k^2|_{\rm min}$ may be proposed.

\vspace{0.3cm}  

\ni {\bf 3. Final remarks}

\vspace{0.3cm}

In our model, stable sterinos are candidates for thermal cold dark matter [1] displaying experimentally the relic abundance $\Omega_{\rm DM}\,h^2 \simeq 0.11$ [6]. If we take the sterino-pair annihilation channels $\bar{\psi} \psi \rightarrow \varphi_{\rm ph}\gamma$ and $\bar{\psi} \psi \rightarrow \bar{f} f$ with $f = e^-, \mu^- , \tau^-, u , d ,c , s , t , b$ (or without $t$ when $m_\psi \ll m_t$) as dominating over the total sterino-pair annihilation cross-section $\sigma_{\rm ann}(\bar{\psi} \psi)$, and require that  $\!<\!\!\sigma_{\rm ann}(\bar{\psi} \psi) 2v_\psi\!\!> \sim $ pbarn (pbarn $\! = \!10^{-36}{\rm cm}^2$) in consistency with $\Omega_{\rm DM}\,h^2 \simeq 0.11$ ({\it i.e.}, that sterinos are WIMP's), we estimate 

\begin{equation} 
m_\psi \sim (13\;\;{\rm to}\;\; 770)\;{\rm GeV}
\end{equation}

\ni under the tentative assumption that 

\begin{equation} 
m^2_\psi \sim (10^{-3}\;\;{\rm to}\;\; 1) <\!\varphi\!>^{\!2}_{\rm vac} \sim m^2_\varphi\;\;,\;\,\, M^2 \sim <\!\varphi\!>^2_{\rm vac} \,,
\end{equation}

\ni and, boldly,

\begin{equation}
f \sim e^2 = 4\pi \alpha \simeq 0.0917 \;\;\;,\;\;\; \zeta \sim 1 
\end{equation}

\ni (for the method, {\it cf.} the third Ref. [1]). This gives

\begin{equation}
M \sim (400\;\;{\rm to}\;\; 770)\;{\rm GeV}\,.
\end{equation}

\ni Here, masses of active leptons and quarks are neglected {\it versus} $E_\psi \sim m_\psi$. 

Then, $\!$we can estimate the nonsingular part of proton elastic cross-section  $\sigma(p\psi\!\rightarrow\!p'\psi')$ given in Eq. (47) as 

\begin{eqnarray}
\sigma_{\rm mag}(p \psi \!\rightarrow p' \!\psi') \!+\! \sigma_{\rm elmag}(p \psi \!\rightarrow \!p' \psi')\!\! &\!\! = \!\!& \!\frac{1}{8\pi}\left(\!\frac{e f \zeta\!<\!\varphi\!>_{\rm vac}}{M^2}\!\right)^{\!2}\!\!\! \frac{m^2_\psi}{(m_p \!+\! m_\psi)^2}\! \left[\!\left(\!\frac{g_p}{2}-\!1\!\right)^2\!\!\!+\!2\left(\!\frac{g_p}{2}\!-\!1\!\right)\!\right] \nonumber \\ \!&\!\! \sim\!\! & 
\!\frac{8\pi^2 \alpha^3}{M^2} \left[\!\left(\frac{g_p}{2}-1\right)^2\!+2\left(\frac{g_p}{2}-1\right)\right] \nonumber \\ \!\!& \!\!=\!\! & \!(13\,{\rm to}\,3.5)\!\!\times\! 10^{-\!10\!}\frac{1}{{\rm GeV}^2} = (5.1\,{\rm to}\,1.4)\!\!\times\! 10^{-\!37\!}{\rm cm}^2 
\end{eqnarray}

\ni and, similarly, the neutron elastic cross-section presented in Eqs. (40) with (41) as $\sigma_{\rm mag}(n \psi \!\rightarrow n' \!\psi') \sim (2.7\;{\rm to}\;0.73)\!\times\!10^{-37}{\rm cm}^2$.

This tentative result may seem to be too large in view of the very recent CoGeNT direct detection experiment ([7], see also [4]), suggesting for (spin-independent) nucleon elastic cross-section on dark matter the approximate figure of $\sim 10^{-40}$ cm$^2$, and for dark-matter mass the range 5-10 GeV roughly. Thus, in the case of spin-independent nucleon cross-sections, such a comparison may mean that the tentative interval $10^{-3}$  to 1 in Eq. (49) ought to be extended and/or the bold value (50) for $f$  diminished. Otherwise, it may turn out that 
the spin dependence of nucleons within nuclei cannot be simply averaged in scattering on dark matter and so, nucleon cross-sections be treated as spin-independent. 

At the end, we would like to underline once more the physical meaning of the new weak interaction (1) that, due to the presence of hidden sector, modifies the \SMo electromagnetism, leading to the total electric source current(4). The latter includes the contribution from the hidden sector,

\begin{equation} 
\delta j_\mu = \partial^\nu(\sqrt{\!f\,} \varphi A_{\mu\nu}),
\end{equation}

\ni which has a generalized magnetic character related to its four-divergence form. Note that the interaction Lagrangian (1) with $F_{\mu \nu} = \partial_\mu A_\nu - \partial_\nu A_\mu$ can be rewritten, up to a four-divergence term, as
 
\begin{equation} 
-\delta j_\mu A^\mu - \frac{1}{2}\,\sqrt{\!f\,}\zeta \bar{\psi} \sigma_{\mu\nu} \psi A^{\mu\nu}.
\end{equation}

\ni This weak interaction Lagrangian together with the \SMo electromagnetic interaction $-j_\mu A^\mu$ gives the modified electromagnetic interaction (including the correlated interaction of $\psi$ and $\bar{\psi}$ with $A^{\mu\nu}$):

\begin{equation} 
-(j_\mu +\delta j_\mu)A^\mu - \frac{1}{2}\,\sqrt{\!f\,}\zeta \bar{\psi} \sigma_{\mu\nu} \psi A^{\mu\nu}.
\end{equation}

\ni Here, of course,

\begin{equation} 
\partial^\mu(j_\mu +\delta j_\mu) = 0
\end{equation}

\ni with $\partial^\mu \delta j_\mu = 0$ satisfied identically due to the antisymmetry of $A_{\mu\nu}$. The modified electromagnetic interaction (55), working both in the hidden and \SM sectors, defines and physically explains the mechanism of photonic portal as a collaboration of couplings $j_\mu A^\mu$ and  $\delta j_\mu A^\mu$ ({\it via} $A^\mu$) as well as $\delta j_\mu A^\mu$ and $(\sqrt{\!f\,}\zeta/2) \bar{\psi}\sigma_{\mu\nu} \psi A^{\mu\nu}$ ({\it via} $A^{\mu\nu}$).  

The mechanism of photonic portal to hidden sector, described in this note, may be embedded as a leading phenomenon in a more extended new weak interaction displaying electroweak symmetry, spontaneously broken by the \SMo Higgs coupling ({\it cf.} the fourth Ref. [1]).

\vspace{1.5cm}

{\centerline{\bf References}}

\vspace{0.4cm}

\baselineskip 0.7cm

{\everypar={\hangindent=0.65truecm}
\parindent=0pt\frenchspacing

{\everypar={\hangindent=0.65truecm}
\parindent=0pt\frenchspacing

~[1]~W.~Kr\'{o}likowski, arXiv: 0803.2977 [{\tt hep--ph}]; {\it Acta Phys. Polon.} {\bf B 39}, 1881 (2008); {\it Acta Phys. Polon.} {\bf B 40}, 111 (2009); {\it Acta Phys. Polon.} {\bf B 40}, 2767 (2009).

\vspace{0.2cm}

~[2]~W.~Kr\'{o}likowski, arXiv: 0909.2498 [{\tt hep--ph}]; arXiv: 0911.5614 [{\tt hep--ph}].

\vspace{0.2cm}

~[3]~{\it Cf. e.g.}  J. March-Russell, S.M. West, D. Cumberbath and D.~Hooper, {\it J. High Energy Phys.} {\bf 0807}, 058 (2008).

\vspace{0.2cm}

~[4]~For some recent calculations {\it cf.}, A.L. Fitzpatrick, D. Hooper and K.M.~Zurek, arXiv: 1003.0014 [{\tt hep--ph}]; S.~Andreas, C.~Arina, T.~Hambye, F-S.~Ling and M.H.G.~Tytgat, arXiv: 1003.2595 [{\tt hep--ph}]; and references therein. 

\vspace{0.2cm}

~[5]~For a related classification of dark matter candidates {\it cf}. P.~Agrawal, Z.~Chacko, C.~Kilic and R.K.~Mishra, arXiv: 1003.1912 [{\tt hep-ph}]; and references therein.

\vspace{0.2cm}

~[6]~C. Amsler {\it et al.} (Particle Data Group), {\it Review of Particle Physics}, {\it Phys. Lett.} {\bf B 667}, 1 (2008). 

\vspace{0.2cm}

~[7]~C.E.~Aalseth {\it et al.} (CoGeNT Collaboration), arXiv: 1002.4703 [{\tt astro-ph.CO}].

\vspace{0.2cm}

\vfill\eject

\end{document}